\documentstyle[preprint,aps]{revtex}	
\begin{document}
\draft
\tighten
%
%
\preprint{
\font\fortssbx=cmssbx10 scaled \magstep2
\hbox to \hsize{
\includegraphics{uwlogo.ps}
\hskip.5in \raise.1in\hbox{\fortssbx University of Wisconsin - Madison}
\hfill$\vtop{\hbox{\bf MAD/PH/882}
                \hbox{\bf LNF-95/015(P)\\}
                \hbox{\bf hep-ph/9503474\\}
                \hbox{March 1995}}$ }
}

%

\title{Tau Decays into Kaons%
\footnote{
We wish to dedicate this work to Roger Decker whose death at an early
age is a great loss to his friends and to the physics community.}
}
\author{Markus Finkemeier}
\address{
INFN-Laboratori Nazionali di Frascati, P.O. Box 13,
00044 Frascati (Roma), Italy}
\author{Erwin Mirkes}
\address{
Department of Physics, University of Wisconsin, Madison, WI 56706, USA}
\maketitle
\begin{abstract}
Predictions for semi-leptonic decay rates of the $\tau$ lepton into
two meson final states
$K^-\pi^0\nu_\tau,\, \overline{K^0}\pi^-\nu_\tau,\, K^0 K^-\nu_\tau,$
and three meson final states
$K^-\pi^-K^+\nu_\tau,$\,    $K^0\pi^-\overline{K^0}\nu_\tau,$\,
$ K_S\pi^- K_S \nu_\tau,$\, $K_S\pi^-K_L      \nu_\tau,$\,
$ K_L\pi^- K_L \nu_\tau,$\, $K^-\pi^0 K^0 \nu_\tau,$\,
$\pi^0\pi^0K^- \nu_\tau,$\, $ K^-\pi^-\pi^+\nu_\tau,$\,
$\pi^-\overline{K^0}\pi^0 \nu_\tau$ are derived.
The hadronic matrix elements are expressed in terms of form factors,
which can be predicted by chiral Lagrangians supplemented by
informations about all possible low-lying resonances in the
different channels. Isospin symmetry relations among the
different final states are carefully taken into account.
The calculated branching ratios are  compared with measured decay
rates where data are available.
\end{abstract}
\newpage
%
%
\narrowtext
\everymath={\displaystyle}

\section{INTRODUCTION}
With the experimental progress in $\tau$ decays an
ideal tool for studying
strong interaction physics has been developed.
In particular, final states with kaons provide a powerful
probe of the strange sector of the weak charged current.

The decay rate for the simplest decay mode with one kaon
$\tau\rightarrow K \nu_{\tau}$ is well
predicted by the the kaon decay constant $f_K$, which can be measured
in $\Gamma(K\rightarrow \mu \bar{\nu}_\mu)$.
Radiative corrections to this decay mode are also available
\cite{markus1}.
The decay rate for
$\tau\rightarrow K\pi\nu_\tau$
on the other hand is the Cabibbo suppressed
analogy to $\tau\rightarrow \rho(\rightarrow \pi\pi) \nu_\tau$,
{\it i.e.} the $K\pi$ system is expected to be dominated by
the $K^\star(892)$ resonance \cite{GR}.
However, the results for this decay rate are fairly sensitive
towards the details of the parameterization of the $K^\star$ resonance,
such as contributions from higher radial excitations, and we
will discuss this dependence in some  detail in Sec.~\ref{h2}.
Since a reliable parameterization of the $K^\star$ propagator
is also needed for the three meson decay modes discussed below,
the informations from the $\tau\rightarrow K\pi\nu_\tau$
decay rates are used to put constraints on the
$K^\star$ resonance  parameterization.

The hadronic matrix elements for three meson final states
with pions and kaons have a much richer structure.
A general parameterization of the form factors in these decay
modes was proposed in \cite{Dec93}.
The physical idea behind this model can be summarized to:
\begin{itemize}
\item In the chiral limit the form factors are normalized to the
      $SU(3)_L\times SU(3)_R$ chiral model.
\item Meson-vertices are independent of momentum.
\item The full momentum dependence is given by normalized
      Breit-Wigner propagators of the resonances occurring in the
      different channels. Resonances occur either in $Q^2$
      [the total invariant hadronic mass squared] which are
      the three body resonances or in the Dalitz plot variables
      $s_i$ which are two body resonances.
\end{itemize}
Parameterizations of the amplitude for
the $3\pi$ final states within this
model can be found in \cite{KueSa,KueMi1,KueMi2}.
In this case the vector form factor is absent due to the
$G$ parity of the pions. The pion decay modes offer a unique
tool for the study of $\rho,\rho'$ resonance parameters in
different hadronic environments, competing well with low energy $e^+e^-$
colliders with energies in the region below 1.7 GeV.
The decay modes involving pions and kaons allow
for axial and vector current contributions at the same time
\cite{braaten,Dec93}.
The vector form factor is related
to the Wess-Zumino anomaly \cite{WZ,kramer}
whereas the axial-vector form factors are predicted by chiral Lagrangians
as mentioned before.
A parameterization of the $\tau^-\rightarrow K^-\pi^-K^+\nu_\tau$
decay mode within this framework has been developed in
\cite{Gom90}. However, our result for this decay mode differs from
the their result (see Sec.~\ref{numeric}).

In the present paper, we reexamine the model used in \cite{Dec93}
for the $\tau$ decay modes involving pions and kaons.
We take into account possible isospin and $SU(3)$ symmetry relations,
which allow for additional resonance contributions to the form factors.
We reanalyze the issue of the strange axial resonances ($K_1$ states) in view
of new experimental results, and we include $\omega-\Phi$ mixing.
The new parameterization leads to sizable differences in the
predictions of the decay rates compared to \cite{Dec93}.
In addition, we derive a parameterization for the final states with two
neutral kaons
$\tau^-\rightarrow K_S\pi^-K_S\nu_\tau, \,
 \tau^-\rightarrow K_L\pi^-K_L\nu_\tau, $ and
$\tau^-\rightarrow K_S\pi^-K_L\nu_\tau$.

The paper is organized as follows:
In Sec.~II we discuss the two meson decay modes
$K^-\pi^0\nu_\tau$,  $\overline{K^0}\pi^-\nu_\tau$ and
$K^0 K^-\nu_\tau$ and fix the parameters of the $K^\star$
resonance.
In Sec.~III we review the general structure of the matrix elements of
the weak hadronic current in the three meson case.
In Sec.~IV (Sec.~V) we discuss in detail the
individual matrix elements of the axial-vector
(vector) current, carefully exploiting isospin and flavour symmetry.
Final states with two neutral kaons are considered in Sec.~VI.
We give our numerical results in Sec.~VII, and finally in Sec.~VIII we
give a brief summary.
%

\section{THE TWO MESON DECAY MODES}
\label{h2}

As mentioned before,
a reliable parameterization of the $K^\star$ propagator
is  needed for the three meson decay modes with kaons.
Since the decay modes
$\tau \rightarrow  K^-\pi^0\nu_\tau$ and
$\tau \rightarrow  \overline{K^0}\pi^-\nu_\tau$
are dominated by the $K^\star$ resonance, we use the
experimental informations on the decay rates into these
final states to fix the parameters of the $K^\star$ resonance.

The  matrix element for the semi-leptonic decay into two mesons
$h_a$ and $h_b$
\begin{equation}
\tau(l,s)\rightarrow\nu(l^{\prime},s^{\prime})
+h_{a}(q_{1},m_{1})+h_{b}(q_{2},m_{2}) \>,
\label{process2h}
\end{equation}
can be expressed in terms of a leptonic ($M_\mu$) and a
hadronic vector  current  ($J^\mu$) as
\begin{equation}
{\cal{M}}=\frac{G}{\sqrt{2}}\,
\bigl(^{\cos\theta_{c}}_{\sin\theta_{c}}\bigr)
\,M_{\mu}J^{\mu} \>.
\label{mdef2h}
\end{equation}
In Eq.~(\ref{mdef2h}),
$G$ denotes the Fermi-coupling constant and  $\theta_c$ is the
Cabibbo angle.
The leptonic  and hadronic  currents are given by
\begin{equation}
M_{\mu}=
\bar{u}(l^{\prime},s^{\prime})\gamma_{\mu}(1-\gamma_{5})u(l,s) \>,
\label{leptoncurrent}
\end{equation}
and
\begin{equation}
J^{\mu}(q_{1},q_{2})=\langle h_{a}(q_{1})h_{b}(q_2)
|V^{\mu}(0)|0\rangle \>.
\label{hadmat2h}
\end{equation}
The hadronic matrix elements for
the Cabibbo suppressed decay modes
$K^-\pi^0\nu_\tau,\, \overline{K^0}\pi^-\nu_\tau$
are dominated by the  $K^\star$ resonance $T_{K^\star}^{(1)}(Q^2)$,
whereas the one for the Cabibbo allowed mode $K^0 K^- $
is dominated by
the high energy tail of the $\rho$.
One has
\begin{eqnarray}
   <  K^-(q_1) \pi^0(q_2) | V^\mu | 0 >
   &=& \frac{1}{\sqrt{2}} <  \overline{K^0}(q_1) \pi^-(q_2)| V^\mu |0>\>,
\nonumber \\
   &=& \frac{1}{\sqrt{2}} T_{K^\star}^{(1)}(Q^2)
    (q_1 - q_2)_\nu\,T^{\mu\nu}
   \label{had2h} \\[2mm] \nonumber
< K^0(q_1) K^-(q_2) | V^\mu | 0 >
  & = & T_\rho^{(1)} (q_1 - q_2)_\nu \,T^{\mu\nu} \>,
\nonumber
\end{eqnarray}
where $Q = q_1 + q_2$ and
$T^{\mu\nu}$ denotes the transverse projector, defined by
\begin{equation}
T^{\mu\nu}=  g^{\mu \nu} - \frac{Q^\mu Q^\nu}{Q^2}  \>.
\label{trans}
\end{equation}
The form and the  normalization
of the hadronic matrix elements in Eq.~(\ref{had2h})
are  fixed  by  chiral symmetry constraints, which determines
the   matrix elements in  the limit of soft
meson momenta $Q^2\rightarrow 0$.
The strong interaction effects beyond the
low energy limit are taken into account by the
vector resonance factors $T_{\rho}^{(1)}(Q^2)$
and $T_{K^\star}^{(1)}(Q^2)$
with the requirement $T_{X}^{(1)}(Q^2=0)=1$ ($X = \rho$, $K^\star$).
Note that in the case of the $K^\star$ we have neglected a scalar
contribution proportional to $Q^\mu$. This scalar part is proportional
to the off-shellness $(m_{K*}^2 - Q^2)$ of the $K^\star$ and therefore
strongly suppressed. We have checked its size numerically and found it to
be negligible.


Note that our results for the hadronic matrix elements for $K^- \pi^0$ and
$\bar{K^0} \pi^-$ in Eq.~(\ref{had2h}) differ from the results in
the Tauola Monte Carlo \cite{tauola}
by an overall factor. In fact we believe that Tauola is off from the correct
normalization by a factor of $2/\sqrt{3}$,
{\it i.e.} the corresponding
matrix elements in Tauola should be multiplied my $\sqrt{3}/2$.
Note furthermore that Tauola does
not include the $K^*(1410)$ in the $K \pi$ resonance, {\it i.e.}
it corresponds to
$\beta_{K^\star} = 0$ in Eq.~(\ref{betakst}).
These two differences with our parameterization almost
cancel each other, such that Tauola (with $\beta_{K^\star} = 0$) gives a number
for the $\tau \to K^\star \nu_\tau$ branching ratio which is very close to our
result with $\beta_{K^\star} = -0.135$ [see below].

If not stated otherwise, we use the
following  form for the two particle
Breit-Wigner propagators
with an energy dependent width $\Gamma_X(s)$ throughout this paper:
\begin{equation}
\mbox{BW}_{X}[s]\equiv {M^2_X\over [M^2_X-s-i\sqrt s \Gamma_X(s)]}\>,
\end{equation}
where $X$ stands for the various resonances of the two meson
channels. For a $1 \to 2$ decay,
the energy dependent width is \cite{Dec93}
\begin{eqnarray*}
\Gamma_X(s)&=& \Gamma_X {M^2_X\over s} \biggl({ p\over p_X}
\biggr)^{2n+1}\,\>, \\[2mm]
 p&=&{1\over 2\sqrt s} \sqrt{(s-(M_1+M_2)^2)(s-(M_1-M_2)^2)}\>,\\[2mm]
p_X&=& {1\over 2M_X}\sqrt{(M^2_X-(M_1+M_2)^2)(M^2_X-(M_1-M_2)^2)} \>,
\end{eqnarray*}
where $n$  is the power of $|p|$ in the matrix element,
{\it i.e.} $n=1$ for the decay modes in this paper.

We use the following parameterizations for the $\rho$
resonance:
\begin{eqnarray}
   T_\rho^{(1)}(s) & = & \frac{1}{1 + \beta_\rho}
  \Big[ \mbox{BW}_\rho(s) + \beta_\rho\, \mbox{BW}_{\rho'}(s)
  \Big] \>,
   \label{beta}
\end{eqnarray}
where
\begin{eqnarray}
\beta_\rho = -0.145\>, & & \nonumber \\
m_\rho = 0.773 \, \mbox{GeV}\>, & & \Gamma_\rho = 0.145 \, \mbox{GeV}\>,
\nonumber \\
m_{\rho'} = 1.370 \, \mbox{GeV}\>, & & \Gamma_{\rho'} = 0.510 \, \mbox{GeV}\>.
 \label{eqnrho}
\end{eqnarray}
These are the values which have been determined from $e^+ e^- \to \pi^+
\pi^-$ in \cite{KueSa} and have been used in \cite{Dec93} for the
non-strange case.
For the vector resonances with strangeness, only the
$K^\star(892)$ was considered in \cite{Dec93}.
Our parameterization for $T_{K^\star}^{(1)}(Q^2)$ allows
for a contribution of the first excitation  ${K^\star}'(1410)$
in analogy to Eq.~(\ref{beta}):
\begin{eqnarray}
   T_{K^\star}^{(1)}(s) & = &\frac{1}{1 + \beta_{K^\star}}
  \Big[ \mbox{BW}_{K^\star}(s) + \beta_{K^\star}\, \mbox{BW}_{{K^\star}'}(s)
  \Big]\>,
  \label{betakst}
\end{eqnarray}
where
\begin{eqnarray}
m_{{K^\star}} = 0.892 \, \mbox{GeV}\>, & &
\Gamma_{{K^\star}} = \, 0.050 \mbox{GeV}\>,
\nonumber \\
m_{{K^\star}'} = 1.412\, \mbox{GeV}\>, & &
\Gamma_{{K^\star}'} = 0.227\, \mbox{GeV}\>.
\label{ksdef}
\end{eqnarray}
In the limit of $SU(3)$ flavour symmetry,
one would expect  a contribution
to the $K^\star$ resonance
from  the first excitation ${K^\star}'(1410)$ with
the  same relative strength  $\beta_{K^\star}=-0.145$
as measured in the non-strange case.
The numerical results for the decay rates
$K^-\pi^0\nu_\tau,\, \overline{K^0}\pi^-\nu_\tau$
are very sensitive to the parameter $\beta_{K^\star}$
in Eq.~(\ref{betakst}).
Since a reliable parameterization of the $K^\star$ propagator will be needed
for the decay modes considered in this paper,
we will use the decay mode $\tau \to K^\star \nu_\tau$
to fix the parameter $\beta_{K^\star}$.

Adding up the two charge modes in Eq.~(\ref{had2h})
\begin{equation}
{{\cal B}(K^\star\nu_\tau) = {\cal B}(K^-  \pi^0 \nu_\tau ) +
{\cal B}(\overline{K^0} \pi^-  \nu_\tau)
}  \>,
\end{equation}
we obtain
\begin{eqnarray}
{\cal B}(K^\star\nu_\tau) =
                  1.00\% && \hspace{1cm}\mbox{for} \hspace{1cm}
                                 \beta_{K^\star} = \,\,\,0. \>,
     \nonumber\\
{\cal B}(K^\star\nu_\tau) =
                  1.28\% && \hspace{1cm}\mbox{for} \hspace{1cm}
                                 \beta_{K^\star} = -0.11 \>,
       \nonumber\\
{\cal B}(K^\star\nu_\tau) =
                  1.36\% && \hspace{1cm}\mbox{for} \hspace{1cm}
                                 \beta_{K^\star} = -0.135 \>,
       \nonumber\\
{\cal B}(K^\star\nu_\tau) =
                  1.44\% && \hspace{1cm}\mbox{for} \hspace{1cm}
                                 \beta_{K^\star} = -0.157 \>.
 \label{eqnkst}
\end{eqnarray}
The main effect is due to the normalization factor
$1/(1+\beta_{K^\star})^2$
multiplying the $K^\star(892)$ contribution, whereas the $K^\star(1410)$
contribution is strongly phase space suppressed.

The latest experimental result on the branching fraction
${\cal B}(K^\star\nu_\tau)$
is $1.36\pm 0.08$ \% \cite{Hel94}.
Thus our results favour a negative value of $\beta_{K^\star}$,
and we obtain
$\beta_{K^\star}=-0.135\pm 0.025$ as a result of this analysis.
This value is remarkably close to the strength of  the
$\rho'$ contribution  to the $\rho$ Breit-Wigner in
Eqs.~(\ref{beta},\ref{eqnrho}),
supporting the use of approximate
$SU(3)$ flavour symmetry.

Note that we use here the values of \cite{RPP94} for the mass and widths
 parameters of
the ${K^\star}'$ in Eqs.~(\ref{betakst},\ref{ksdef}),
whereas the $\rho'$ mass
and width parameters in Eq.~(\ref{eqnrho}) have
been determined from a fit to the $Q^2$ distribution in $\tau\to
2\pi\nu_\tau$ \cite{KueSa}.
Indeed, a more reliable determination of the parameters of the off-shell
$K^\star$ propagator (including $\beta_{K^\star}$)
could be obtained from a fit to the $Q^2$ distribution
in $\tau\to K^\star \nu_\tau$.
Unfortunately, there is no experimental information of the  $Q^2$
dependence in this decay mode available, and  we use therefore the simpler
approach of fixing $\beta_{K^\star}$
from the branching ratio and using the
${K^\star}'$ parameters from \cite{RPP94}.

For the decay into two kaons we obtain from the matrix element in
Eq.~(\ref{had2h})
\begin{equation}
   {\cal B}(K^0 K^- \nu_\tau) = 0.11 \%\>,
\end{equation}
in good agreement with the recent world average
$   {\cal B}(K^0 K^- \nu_\tau) = 0.13 \pm 0.04 \% $ \cite{Hel94}.

\section{GENERAL STRUCTURE OF THE WEAK MATRIX ELEMENTS IN THE
         THREE MESON MODES}

Let us briefly recapitulate the general structure of the semi-leptonic
decay
\begin{equation}
\tau(l,s)\rightarrow\nu(l^{\prime},s^{\prime})
+h_{a}(q_{1},m_{1})+h_{b}(q_{2},m_{2})+h_{c}(q_{3},m_{3}) \>,
\label{process}
\end{equation}
as introduced in  \cite{KueMi1}.
In Eq.~(\ref{process}),  $h(q_i,m_i)$ are pseudoscalar mesons
and in our case of interest, at least one of the mesons is a kaon.
The  matrix element is
\begin{equation}
{\cal{M}}=\frac{G}{\sqrt{2}}\,
\bigl(^{\cos\theta_{c}}_{\sin\theta_{c}}\bigr)
\,M_{\mu}J^{\mu} \>,
\label{mdef}
\end{equation}
%
%
where the  cosine and the sine of the Cabibbo angle ($\theta_C$)
in Eq.~(\ref{mdef}) have to be used for Cabibbo allowed $\Delta S=0$
[{\it i.e.} final states with two kaons] and
Cabibbo suppressed $|\Delta S|=1$ [{\it i.e.} final states with one kaon]
decays, respectively.
The leptonic ($M_\mu$) current
is already given in Eq.~(\ref{leptoncurrent}) and the
hadronic ($J^\mu$) currents can be written as
\begin{equation}
J^{\mu}(q_{1},q_{2},q_{3})=\langle h_{a}(q_{1})h_{b}(q_{2})h_{c}(q_{3})
|V^{\mu}(0)-A^{\mu}(0)|0\rangle \>.
\label{hadmat}
\end{equation}
$V^\mu$ and $A^\mu$ are the vector and axial-vector quark currents,
respectively.
The most general ansatz for the matrix element of the
quark current $J^{\mu}$  in Eq.~(\ref{hadmat})
is characterized by four form factors
\begin{eqnarray}
J^{\mu}(q_{1},q_{2},q_{3})
&=&   V_{1}^{\mu}\,F_{1}
    + V_{2}^{\mu}\,F_{2}
    +\,i\, V_{3}^{\mu}\,F_{3}
    + V_{4}^{\mu}\,F_{4} \>,
    \label{f1234}
\end{eqnarray}
with
\begin{equation}
\begin{array}{ll}
V_{1}^{\mu}&= (q_{1}-q_{3})_{\nu}\,T^{\mu\nu}  \>,\\[2mm]
V_{2}^{\mu}&= (q_{2}-q_{3})_{\nu}\,T^{\mu\nu}  \>,\\[2mm]
V_{3}^{\mu}&= \epsilon^{\mu\alpha\beta\gamma}q_{1\,\alpha}q_{2\,\beta}
                                             q_{3\,\gamma} \>,
\\[2mm]
V_{4}^{\mu}&=q_{1}^{\mu}+q_{2}^{\mu}+q_{3}^{\mu}\,=Q^{\mu} \>.
    \end{array}
\label{videf}
\end{equation}
$T^{\mu\nu}$ denotes the transverse projector, defined in Eq.~(\ref{trans}).
Since the strong interaction conserves parity, the axial-vector
current induces the form factors $F_1, F_2$ and $F_4$ while the vector
current induces $F_3$.
In the limit of vanishing quark masses, the weak axial-vector
current is conserved and this implies that the scalar form factor
$F_4$ vanishes. The massive pseudoscalars  give a
contribution to $F_4$, however, the effect is very small
\cite{dfm} and we will neglect this contribution in the subsequent discussion,
{\it i.e.} we set $F_4$ equal  to zero.
All form factors $F_i$ are in general functions of $Q^2$,
$s_1=(q_2+q_3)^2, s_2=(q_1+q_3)^2$ and $s_3=(q_1+q_2)^2$.

A specific model for the form factors $F_i$ for various three meson final
states was derived in \cite{Dec93}.
This model takes into
account the chiral symmetry constraints of QCD as well as the
resonance phenomena present in $\tau$ decays and has been
developed in \cite{chiral}.
Let us briefly summarize the results (for more details see \cite{Dec93}):
The chiral symmetry constraints lead to the following expression for
the hadronic axial-vector current in the limit of soft meson momenta:
\begin{eqnarray}
\langle h_a h_b h_c|A^\mu|0\rangle &=&
\frac{2\sqrt{2} A^{(abc)}}{3f_\pi}
\left\{ G^{(abc)}_{1,soft} (q_1-q_3)_\nu +
G^{(abc)}_{2,soft}(q_2-q_3)_\nu\right\}\,
     \,\,T^{\mu\nu}\>,
\label{m1}
\end{eqnarray}
where the coefficients $A^{(abc)}$ are given in Tab.~\ref{tab1} for the
various decay modes
and the coefficients $G^{(abc)}_{i,soft}$ are all equal to one
\begin{equation}
   G^{(abc)}_{i,soft}  = 1\>,  \hspace{2cm}(i=1,2)
\end{equation}
except for $G_{2,soft}^{(K^- \pi^0 K^0)}$ and
$G_{2,soft}^{(\pi^- \overline{K^0} \pi^0)}$
which vanish:
%
%
$
 G_{2,soft}^{(K^- \pi^0 K^0)} = G_{2,soft}^{(\pi^- \overline{K^0}\pi^0)} = 0\>.
$
%
%
The vector current arises from the Wess-Zumino Lagrangian for the
axial anomaly \cite{WZ,kramer}. One obtains in the low energy limit:
\begin{eqnarray}
\langle h_a h_b h_c|V^\mu|0\rangle &=&
\frac{i}{2\sqrt{2}\pi^2 f_\pi^3} A^{(abc)}
\epsilon^{\mu\alpha\beta\gamma} q_{1\,\alpha}q_{2\,\beta}q_{3\,\gamma}
\, G_{3,soft}^{(abc)} \>.
\label{m2}
\end{eqnarray}
The coefficients $A^{(abc)}$ for the vector current are given in
Tab.~\ref{tab2}
and
%
%
$
G_{3,soft}^{(abc)}=1\>,
$
%
%
except for $G_{3,soft}^{(K^- \pi^0 K^0)}$ and
$G_{3,soft}^{(\pi^0 \pi^0 K^- )}$
which vanish:
%
%
$
  G_{3,soft}^{(K^- \pi^0 K^0)} = G_{3,soft}^{(\pi^0 \pi^0 K^-)} = 0\>.
$

As mentioned before, the strong interaction effects beyond the
low energy limit are taken into account by inserting resonance
form factors $G_{1,2,3}(Q^2,s_i)$ [$i=1,2,3$]
into the amplitudes of Eqs.~(\ref{m1},\ref{m2})
with the requirement $G^{(abc)}_{1,2,3}\rightarrow G^{(abc)}_{1,2,3,soft}= 1 $
 in the limit
$Q^2,\,s_i \to 0$,
except for $G_{2,3}^{(K^- \pi^0 K^0)}$, $G_2^{(\pi^- \overline{K^0} \pi^0)}$
and $G_{3}^{(\pi^0 \pi^0 K^-)}$,
which vanish in the chiral limit.

In fact, the functions $G^{(abc)}_{1,2,3}$ are
products of normalized Breit-Wigner resonances in $Q^2$ and
$s_i$.
Comparing Eqs.~(\ref{m1},\ref{m2}) with the
general expansion in Eqs.~(\ref{f1234},\ref{videf}) leads to
\begin{eqnarray}
F^{(abc)}_{1}(Q^2,s_2,s_3)&=&{2\sqrt 2 A^{(abc)}\over 3f_\pi}
                              G_{1}^{(abc)}(Q^2,s_2,s_3) \>,
                              \label{f1}\\
F^{(abc)}_{2}(Q^2,s_1,s_3)&=&{2\sqrt 2 A^{(abc)}\over 3f_\pi}
                              G_{2}^{(abc)}(Q^2,s_1,s_3) \>,
                              \label{f2}\\
F^{(abc)}_{3}(Q^2,s_1,s_2,s_3) &=& {A^{(abc)}\over 2\sqrt 2\pi^2f^3_\pi}
                              G_3^{(abc)}(Q^2,s_1,s_2,s_3)\>.
                              \label{f3}
\end{eqnarray}
The Breit-Wigner functions $G_{1,2}$ ($G_3$) are listed in
Tab.~\ref{tab1} (\ref{tab2})
for the various decay modes and the precise form will be
discussed in the subsequent sections.
Note that by convenient ordering of the mesons, the two body resonances
in $F_1$ ($F_2$) occur only in the variables $s_2,s_3$ ($s_1,s_3)$.

\section{ISOSPIN AND FLAVOUR SYMMETRY RELATIONS IN THE AXIAL-VECTOR CURRENT}

Amongst the hadronic matrix elements for the hadronic final states
$K^- \pi^- K^+$, $K^0 \pi^-
\overline{K^0}$ and $K^- \pi^0 K^0$
(which are  versions of $K \pi\bar{K}$ with different third
components of the isospin)
and amongst those for $\pi^0 \pi^0 K^-$, $K^- \pi^- \pi^+$ and $\pi^-
\overline{K^0} \pi^0$ (different versions of $\overline{K} \pi \pi$),
there are isospin symmetry relations which have not been fully exploited
in \cite{Dec93}. Consider for example the final state $K^- \pi^- K^+$,
which is dominated by a $(K^\star)^0$ resonance in $\pi^- K^+$.
In Tab.~\ref{tab1} of \cite{Dec93}, this is taken into account by
\begin{equation}
   G_2^{(K^- \pi^- K^+)} =  \mbox{BW}_{A_1}(Q^2) T_{K^\star}(s_1) \>.
\end{equation}
However, by isospin symmetry, the amplitudes $(K^\star)^0 \to
\pi^0 K^0$ and $(K^\star)^0 \to \pi^- K^+$ have a ratio of $\sqrt{1/3} :
(-) \sqrt{2/3}$. Taking into account the different normalization
coefficients $A^{(abc)}$, this immediately leads to
\begin{equation}
  G_2^{(K^- \pi^0 K^0)} = \mbox{BW}_{A_1}(Q^2)
  \left(
  \frac{1}{3} T_{K^\star}(s_1) + \cdots \right) \>,
\end{equation}
which differs from $G_2^{(K^- \pi^0 K^0)} = 0$
in Tab.~\ref{tab1} of \cite{Dec93}. The dots indicate that
there must be an additional contribution which ensures that
$G_2^{(K^- \pi^0 K^0)} \to 0$ for $Q^2$, $s_i \to 0$
as required by the chiral limit (see below).

In fact, by using all the relevant Clebsch-Gordon coefficients,
we can predict the matrix element for $K^- \pi^0
K^0$ from those for $K^- \pi^- K^+$ and $K^0 \pi^- \overline{K^0}$, and
the one for $\pi^- \overline{K^0} \pi^0$ can be predicted from those of
$\pi^0 \pi^0 K^-$ and $K^- \pi^- \pi^+$.
In using the Clebsch-Gordons, some care is needed because in
\cite{Dec93}, the phase conventions are {\em not} identical to the
Condon-Shortley-de Swart ones.
And so, in a first step Tab.~\ref{tab1} of
 \cite{Dec93} has to be transformed into
Condon-Shortley-de Swart by multiplying $A^{(abc)}$ with $(-1)$ for each
 $\pi^-$, $K^-$, $K^0$, $\overline{K^0}$ in the final state. Then the
Clebsch-Gordans may be used, and finally the results are transformed
back into the conventions of \cite{Dec93}. This issue does not affect
the above example, because $< K^- \pi^- K^+|$ and $<K^- \pi^0 K^0|$ are
even under this transformation.

Taking the matrix elements for $K^- \pi^- K^+$ and $K^0 \pi^-
\overline{K^0}$, $\pi^0 \pi^0 K^-$ and $K^- \pi^- \pi^+$ from
\cite{Dec93}, we obtain (see also Tab.~\ref{tab1})
\begin{eqnarray}
  \lefteqn{  < K^- \pi^0 K^0 | A^\mu | 0 >
   =
  \frac{2 \sqrt{2}}{3 f_\pi}
  \frac{3 \cos \theta_c}{2 \sqrt{2}} \mbox{BW}_{A_1}(Q^2)
    \,\,T^{\mu\nu} }
\nonumber \\ & & \times
  \left\{ \Big[ \frac{2}{3} T_\rho^{(1)}(s_2) + \frac{1}{3}
  T_{K^\star}(s_3) \Big] (q_1 - q_3)_\nu
  + \frac{1}{3} \Big[  T_{K^\star}(s_1) -
  T_{K^\star}(s_3) \Big] (q_2 - q_3)_\nu
  \right\} \>,   \\[6mm]
  \lefteqn{  < \pi^- \overline{K^0} \pi^0 | A^\mu | 0 >
  =
  \frac{2 \sqrt{2}}{3 f_\pi}
  \frac{3 \sin \theta_c}{2 \sqrt{2}} \mbox{BW}_{K_1}(Q^2)
    \,\,T^{\mu\nu} }
\nonumber  \\ & & \times
  \left\{ \Big[ \frac{2}{3} T_\rho^{(1)}(s_2) + \frac{1}{3}
  T_{K^\star}(s_3) \Big] (q_1 - q_3)_\nu
  + \frac{1}{3}\Big[  T_{K^\star}(s_1) -
  T_{K^\star}(s_3) \Big] (q_2 - q_3)_\nu  \right\} \>.
\end{eqnarray}
Note that the form factor $G_2$ in the two decay modes
$K^- \pi^0 K^0$ and $\pi^- \overline{K^0} \pi^0$
receive   contributions from $T_{K^\star}$ resonances,
which however vanish in the low energy chiral limit,
where our Breit-Wigner propagators are normalized to one.
In \cite{Dec93}, this contribution was not taken into account.

The parameterizations of the two particle
resonances have been determined
in Sec.~\ref{h2},
where we found $\beta_{K^\star} = - 0.135 \pm 0.025$ for
the $K^\star$ resonance. Our numerical results for the three meson decay
modes are very sensitive to the parameter $\beta_{K^\star}$ and we will
discuss this dependence in more detail in Sec.~\ref{numeric}.

Let us now  discuss the three particle resonances. As in \cite{Dec93},
we use the $A_1$ resonance  in the non-strange case  with energy dependent
width
\begin{equation}
  \mbox{BW}_{A_1}(s) =
  \frac{m_{A_1}^2} {m_{A_1}^2 - s - i m_{A_1} \Gamma_{A_1} g(s) /
  g(m_{A_1})}\>,
\end{equation}
\begin{eqnarray}
m_{A_1}       = 1.251 \,\, \mbox{GeV}\>, & & \hspace{1cm}
\Gamma_{A_1 } = \, 0.475\,\, \mbox{GeV}\>.
\label{a1}
\end{eqnarray}
where the  function $g(s)$ has been calculated in
\cite{KueSa}. Note that we use a smaller $A_1$ width than used
in \cite{Dec93,KueMi1,KueMi2} ($\Gamma_{A_1}=0.599$ GeV).
The formalism of structure functions, which
was developed in \cite{KueMi1,KueMi2}, allow for a much more
detailed test of the hadronic matrix elements and the resonance parameters
than it is possible by a rate
measurement alone. However,
the ratios of structure functions, which were
predicted in \cite{KueMi1,KueMi2}
for the $\tau^-\rightarrow \pi^-\pi^-\pi^+\nu_\tau$
decay mode  and found to be in
good agreement with the experimental data \cite{wermes}, are
not sensitive to the $A_1$ mass and width parameters since they
cancel in the ratios. On the other hand, a measurement of the $Q^2$ dependence
of the structure functions itself would be very sensitive to the $A_1$
 parameters
\cite{KueMi3}
(after having fixed the $\rho$ parameters in the ratios).
An $A_1$ width of 0.475 GeV predicts a branching fraction
${\cal B}(\tau^-\rightarrow \pi^-\pi^-\pi^+\nu_\tau)$ of 8.6 \%
(compared to 6.3 \% for $\Gamma_{A_1}=599$ GeV),
which seems to be favored by the experimental results
for ${\cal B}(3h\nu_\tau)= {\cal B}(\pi^-\pi^-\pi^+\nu_\tau) +
                     {\cal B}( K^- \pi^-\pi^+\nu_\tau) +
                     {\cal B}( K^- \pi^- K^+ \nu_\tau) +
                     {\cal B}( K^-  K^-  K^+ \nu_\tau) = 9.51 \pm 0.18$
\cite{Hel94}.

There are two relevant particles which have the right
quantum numbers in  the case of the three particle resonances with
strangeness,  {\it i.e.} the $K_1(1270)$ and the $K_1(1400)$. Results from
the TPC/Two-Gamma Collaboration had indicated that only the $K_1(1400)$
occurs in $\tau$ decays into $K\pi\pi$ \cite{TPC}. Thus in
\cite{Dec93} only the $K_1(1400)$ was included. Recent experimental
data, however, do not support the results from \cite{TPC}, but rather
indicate that both $K_1$ resonances are produced, with substantially
more $K_1(1270)$ than $K_1(1400)$ \cite{Smi94}. In fact, this is much
more natural within the framework of the resonance enhanced chiral
calculation. There are two types of decay chains which lead to the
$K\pi\pi$ states, {\it i.e.}
\begin{itemize}
\item[a)] $K_1 \to K^\star \pi, K^\star \to K \pi \>,$
\item[b)] $K_1 \to \rho K, \rho \to \pi \pi       \>.$
\end{itemize}
The $K_1(1400)$ decays exclusively into $K^\star \pi$, while the
$K_1(1270)$ decays mainly into $K \rho$, with a smaller branching ratio
into $K^\star \pi$. Thus one should assume that the decay chain $(a)$ is
dominated by the $K_1(1400)$ with a little admixture of $K_1(1270)$,
whereas in (b) only the $K_1(1270)$ contributes:
\begin{eqnarray}
   T_{K_1}^{(a)}(s) & = & \frac{1}{1 + \xi}
   \Big[ \mbox{BW}_{K_1(1400)}(s) + \xi \mbox{BW}_{K_1(1270)} (s)
   \Big] \>,
\nonumber \\[2mm]
   T_{K_1}^{(b)}(s) & = & \mbox{BW}_{K_1(1270)}(s) \>.
\end{eqnarray}
Using the ratio $\Gamma(K_1(1270) \to K^\star \pi) /
\Gamma(K_1(1400) \to K^\star \pi)$, we obtain after phase space
correction:
\begin{equation}
   |\xi| = 0.33.
    \label{xidef}
\end{equation}
Thus $\xi$ is determined up to a sign ambiguity. We will consider both
possibilities below and find that $\xi = -0.33$ leads to decay rates
which strongly disagree with experimental data (see Sec.~\ref{numeric}).
Therefore we believe
that $\xi = + 0.33$ is the physical choice.
In case of the $K_1$ resonances we use  normalized Breit-Wigner propagators
with constant  widths:
\begin{equation}
BW_{K_1}[s]\equiv
\frac{-m^2_{K_1}+im_{K_1}\Gamma_{K_1}}{[s-m^2_{K_1}+im_{K_1}
\Gamma_{K_1}]},
\label{bwc}
\end{equation}
and \cite{RPP94}
\begin{equation}
\begin{array}{ll}
  m_{K_1}(1400) = 1.402 \ {\rm GeV}\>,
& \Gamma_{K_1}(1400)= 0.174 \,{\rm GeV}\>,\\
  m_{K_1}(1270) = 1.270 \ {\rm GeV}\>,
 & \Gamma_{K_1}(1270)= 0.090 {\rm GeV}\>.\\
\end{array}
\end{equation}
All matrix elements of the axial weak current are
summarized in Tab.~\ref{tab1}.
The parameterization for the decay modes with two neutral kaons
will be discussed in Sec.~\ref{neutral}.

\section{ISOSPIN AND FLAVOUR SYMMETRY RELATIONS IN THE VECTOR CURRENT}

We consider now  the vector current with its contributions from the
anomaly.
Let us first correct an error in
Tab.~\ref{tab2} of  \cite{Dec93}. The two kaon resonance in the decay channels
$K^-\pi^- K^+$ and $K^0 \pi^- \overline{K^0}$ can not be a $\rho$
resonance.
Due to $G$ parity conservation, the vector resonance $V$ in the decay chain
$W^- \to \rho^-$, $\rho^- \to V \pi^-$, $V \to (K^- K^+, K^0 \overline{K^0})$
must have $G = -1$. Thus only the $\omega$ and the $\Phi$ qualify.
In the limit of exact $SU(3)$ flavour symmetry, the coupling $\rho \Phi
\pi$ vanishes,
such that only the $\omega$ contributes in this limit.
Taking  $\omega \Phi$ mixing into account, we use
\begin{equation}
   T_\omega(s) = \frac{1}{1 + \epsilon} [ \mbox{BW}_\omega(s) +
   \epsilon \mbox{BW}_\Phi(s) ]\>,
\end{equation}
where \cite{epsi}
\begin{equation}
    \epsilon = 0.05\>.
\end{equation}
Obviously, use of the $\omega$ instead of
the $\rho$ changes the relevant isospin symmetry relations.

We use Breit-Wigner propagators with fixed widths,
of the same form as the one  used for the
$K_1$ resonances in Eq.~(\ref{bwc}), with \cite{RPP94}
\begin{equation}
\begin{array}{ll}
 m_{\omega} = 0.782 \ {\rm GeV}\>, & \Gamma_{\omega}= 0.00843 \,{\rm GeV}\>,\\
 m_{\phi}   = 1.020 \ {\rm GeV}\>, & \Gamma_{\phi}  = 0.00443 \,{\rm GeV}\>.\\
\end{array}
\end{equation}

Similarly to the case of the axial-vector current,
there are also  different decay chains in the vector current
case  which have either $K^\star$ or
$\rho$/$\omega$ two particle resonances. Whereas in the case of the
axial-vector current, the relative strength of the strangeness $S=0$ and $S=-1$
resonances could be fixed by the chiral limit in \cite{Dec93},
an additional free parameter $\alpha$ was
introduced in the case of the vector current,
parameterizing the relative strength of the two body resonances.
In the case of the $K^- \pi^- K^+$ and $K^0 \pi^- \overline{K^0}$
states, the parameterization in \cite{Dec93} was
\begin{equation}
    G_3^{(K^- \pi^- K^+)} =
    G_3^{(K^0 \pi^- \overline{K^0})}
   = T_\rho^{(2)}(Q^2)
  \frac{T_\rho^{(1)} (s_2) + \alpha T_{K^\star}(S_1)} {1 + \alpha}\>.
\end{equation}
where $T_\rho^{(1)}$ should read $T_\omega$, as we have explained above.
However, by writing the relevant vertices
in a flavour invariant way,
\begin{eqnarray}
   g_{VV\pi} \epsilon^{\mu\nu\alpha\beta}
   \mbox{tr}\Big( \partial_\mu V_\nu \partial_\alpha V_\beta \Pi\Big)
\nonumber \\ \nonumber \\
   i g_{V\pi\pi} \mbox{tr} \Big( V_\mu
  [\partial^\mu  \Pi, \Pi ]\Big)
\end{eqnarray}
(where $\Pi = \pi^a \lambda^a$ describes the pseudoscalar mesons, and
similarly $V_\mu$ parameterizes the vector mesons)
the parameter $\alpha$ can be determined, with the result
\begin{equation}
   \alpha = 1/\sqrt{2}\>.
\end{equation}
Furthermore, as in the case of the axial-vector current,
there are isospin symmetry relations between the
matrix elements.

Our final result for the matrix elements of the vector current,
taking into account all isospin and flavour symmetry relations,
can be found in Tab.~\ref{tab2}. Note
that there are two cases in which the
anomaly contribution vanishes in the low energy limit, viz.\ $K^- \pi^0
K^0$ (because ``anomalies do not develop second class currents'', see
also
\cite{kramer}) and $\pi^0 \pi^0 K^-$ (because of Bose symmetry and the
antisymmetry of the anomaly). In both
cases,
however, the presence of resonances leads to non-vanishing
contributions in the higher energy regime. Their precise forms are
predicted from other matrix elements which are non zero in the chiral
limit and to which they are related by isospin symmetry.
These contributions were not taken into account in the
parameterization used in \cite{Dec93}.

Let us now discuss the three particle vector resonances $T_\rho^{(2)}$
and $T_{K^\star}^{(2)}$. In \cite{Dec93}, a form for $T_\rho^{(2)}$
including $\rho$, $\rho'$ and $\rho''$ was used, which was obtained from
a fit to $e^+ e^- \to \eta \pi \pi$ data \cite{DM2,Gom90}. In the three
particle
vector resonance with strangeness, only the $K^\star(892)$ was included.
However, the higher radials will also  be included in this paper and
we will therefore use
\begin{eqnarray}
   T_\rho^{(2)} & = & \frac{1}{1 + \lambda + \mu}
  \Big[ \mbox{BW}_\rho(s) + \lambda \mbox{BW}_{\rho'}(s)
  + \mu \mbox{BW}_{\rho''}(s)
  \Big] \>,
\nonumber \\[2mm]
   T_{K^\star}^{(2)} & = &\frac{1}{1 + \lambda + \mu}
  \Big[ \mbox{BW}_{K^\star}(s) + \lambda \mbox{BW}_{{K^\star}'}(s)
  + \mu \mbox{BW}_{{K^\star}''}(s)
  \Big]\>,
\end{eqnarray}
where
\begin{equation}
  \lambda = \frac{6.5}{-26} = -0.25\>,
  \qquad
   \mu = \frac{1}{-26} = - 0.038 \>.
\end{equation}
For the $\rho$'s,
\begin{eqnarray}
m_\rho = 0.773 \, \mbox{GeV}\>,    & &\Gamma_\rho = 0.145 \, \mbox{GeV}\>,
   \nonumber \\
m_{\rho'} = 1.500 \, \mbox{GeV}\>, & &\Gamma_{\rho'} = 0.220 \, \mbox{GeV}\>,
\nonumber \\
m_{\rho''} = 1.750 \,\mbox{GeV}\>, & &\Gamma_{\rho''} = 0.120 \, \mbox{GeV}\>.
\end{eqnarray}
This is exactly the parameterization of \cite{Gom90}, which was used in
\cite{Dec93}, written in a slightly different way.
For the $K^\star$'s, we use the values \cite{RPP94}:
\begin{eqnarray}
m_{{K^\star}} = 0.892 \, \mbox{GeV}\>, & &
\Gamma_{{K^\star}} = \, 0.050 \mbox{GeV}\>,
\nonumber \\
m_{{K^\star}'} = 1.412\, \mbox{GeV}\>, & &
\Gamma_{{K^\star}'} = 0.227\, \mbox{GeV}\>,
\nonumber \\
m_{{K^\star}''} = 1.714 \, \mbox{GeV}\>, & &
\Gamma_{{K^\star}''} = 0.323 \, \mbox{GeV}\>.
\nonumber \\
\end{eqnarray}
We use energy dependent widths here.

Similar remarks as have been made above for the $T_{K^\star}^{(1)}$
apply
here as well. The parameters $\lambda$ and $\mu$, as well as the $\rho'$
and $\rho''$ parameters of $T_\rho^{(2)}$ have been obtained from a fit
to data. For the $K^\star$ in the anomalous channel,
we use the same parameters $\lambda$ and $\mu$ as for the $\rho$ with
${K^\star}'$, ${K^\star}''$ parameters taken from \cite{RPP94}.
In principle, a more reliable
determination of $T_{K^\star}^{(2)}$
could be obtained from a fit to suitable data.
It should be noted, however, that the numerical significance of these
details is fairly small, because of the small vector channel
contribution to the relevant decay modes
(see the last 3 entries in column 3 of Tab.~\ref{tab3}).

\section{FINAL STATES WITH TWO NEUTRAL KAONS}
\label{neutral}

In \cite{Dec93} the hadronic matrix elements have been expressed in
terms of the strong interaction eigenstates $K^0$ and $\overline{K^0}$.
The actual measurements, however, are performed in terms of the weak
interaction eigenstates $K_S$ and $K_L$. Neglecting $CP$ violation, the
relation between these are given by
\begin{eqnarray}
   K_S = \frac{K^0 - \overline{K^0}}{\sqrt{2}}
& \qquad &
   K_L = \frac{K^0 + \overline{K^0}}{\sqrt{2}}
\nonumber \\ \nonumber \\
   K_0 = \frac{K_L + K_S}{\sqrt{2}}
& \qquad &
   \overline{K_0} = \frac{K_L - K_S}{\sqrt{2}}
\label{klsdef}
\end{eqnarray}
Since the weak current $J^\mu$ produces states with $|S|=0$, $1$ only,
{\it i.e.}
\begin{equation}
   <K^0 \pi^- {K^0} | J^\mu | 0 >
   = <\overline{K^0} \pi^- \overline{K^0} | J^\mu | 0 > = 0
\label{k00}
\end{equation}
we can easily show that
\begin{eqnarray} \label{eqn1}
\lefteqn{   <K_L(q_1) \pi^-(q_2) K_L(q_3) | J^\mu | 0 >
 = -  <K_S(q_1) \pi^-(q_2) K_S(q_3) | J^\mu | 0 >}
\nonumber \\
 & = &\frac{1}{2} \left\{ < K^0(q_1) \pi^- (q_2) \overline{K^0}(q_3)
  | J^\mu | 0>
  +  < \overline{K^0}(q_1) \pi^- (q_2) {K^0}(q_3) | J^\mu | 0>
  \right\} \>,
\nonumber \\ \nonumber \\
\lefteqn{   <K_S(q_1) \pi^-(q_2) K_L(q_3) | J^\mu | 0 >}
\nonumber \\
 & = & \frac{1}{2} \left\{ < K^0(q_1) \pi^- (q_2) \overline{K^0}(q_3)
  | J^\mu | 0>
  -  < \overline{K^0}(q_1) \pi^- (q_2) {K^0}(q_3) | J^\mu | 0>
  \right\}\>.
\end{eqnarray}
Let  us start with the discussion of the axial part of the weak current. From
Eq.~(\ref{m1}) and Tab.~\ref{tab1}, we have
\begin{eqnarray}
\lefteqn{  <K^0 \pi^- \overline{K^0} | A^\mu | 0 >
  = \frac{2 \sqrt{2}}{3 f_\pi} \left(\frac{- \cos \theta_c}{2}\right)
   \mbox{BW}_{A_1}(Q^2)\hspace{3cm} }
\nonumber \\[2mm]
& &
   \times \left\{ T_\rho^{(1)}(s_2) (q_1 - q_3)_\nu
  + T_{K^\star}(s_1) (q_2 - q_3)_\nu \right\} \,\,T^{\mu\nu}\>.
\end{eqnarray}
Using Eq.~(\ref{eqn1}) one obtains
\begin{eqnarray}
\lefteqn{   <K_L(q_1) \pi^-(q_2) K_L(q_3) | A^\mu | 0 >
 = -  <K_S(q_1) \pi^-(q_2) K_S(q_3) | A^\mu | 0 > = }
\nonumber \\[2mm]
 &  & \frac{2 \sqrt{2}}{3 f_\pi} \frac{\cos \theta_C}{4}
   \mbox{BW}_{A_1}(Q^2)
  \left\{ T_{K^\star}^{(1)}(s_3) (q_1 - q_3)_\nu
   - \Big[ T_{K^\star}^{(1)}(s_1) + T_{K^\star}^{(1)}(s_3) \Big]
    (q_2 - q_3)_\nu \right\}\,\,T^{\mu\nu}\>,
 \nonumber \\[4mm]
\lefteqn{   <K_S(q_1) \pi^-(q_2) K_L(q_3) | A^\mu | 0 >
 =  \frac{2 \sqrt{2}}{3 f_\pi} \left(\frac{- \cos \theta_C}{4}\right)
   \mbox{BW}_{A_1}(Q^2) }
\nonumber \\[2mm] & & \times
  \left\{ \Big[ 2 T_\rho^{(1)}(s_2) + T_{K^\star}^{(1)}(s_3) \Big]
  (q_1 - q_3)_\nu
   + \Big[ T_{K^\star}^{(1)}(s_1) - T_{K^\star}^{(1)}(s_3) \Big]
    (q_2 - q_3)_\nu \right\}\,\,T^{\mu\nu}\>.
\end{eqnarray}
We turn now to  the matrix element of the vector current.
According to 
Tab.~\ref{tab1}, the
relevant matrix element is
\begin{equation}
 <K^0 \pi^- \overline{K^0} | V^\mu | 0 >
 =
 \frac{i}{2 \sqrt{2} \pi^2 f_\pi^3} \cos \theta_c
  T_\rho^{(2)} (Q^2)
  (\sqrt{2} - 1) \big[
  \sqrt{2} T_\omega (s_2) +  T_{K^\star}^{(1)}(s_1) \big]
  \epsilon^{\mu\alpha\beta\gamma} q_{1\alpha} q_{2 \beta} q_{3 \gamma}\>.
\end{equation}
In this case Eq.~(\ref{eqn1}) yields (see Tab.~\ref{tab2})
\begin{eqnarray}
\lefteqn{   <K_L \pi^- K_L | V^\mu | 0 > =
   -   <K_S \pi^- K_S | V^\mu | 0 > }
\nonumber \\
& = & \frac{i}{2 \sqrt{2} \pi^2 f_\pi^3}
  \frac{\cos \theta_c}{2}
  (\sqrt{2} - 1)
  T_\rho^{(2)} (Q^2) \Big[ T_{K^\star}(s_1) - T_{K^\star}(s_3)
  \Big]
  \epsilon^{{\mu\alpha\beta\gamma}} q_{1\alpha} q_{2\beta} q_{3 \gamma}\>,
\nonumber \\[4mm]
\lefteqn{   <K_S \pi^- K_L | V^\mu | 0 > }
\nonumber \\
& = & \frac{i}{2 \sqrt{2} \pi^2 f_\pi^3}
  \frac{\cos \theta_c}{2}
  (\sqrt{2} - 1)
  T_\rho^{(2)} (Q^2) \Big[2 \sqrt{2} T_{\omega}(s_2)
  + T_{K^\star}^{(1)}(s_1) + T_{K^\star}^{(1)}(s_3)
   \Big]
  \epsilon^{{\mu\alpha\beta\gamma}} q_{1\alpha} q_{2\beta} q_{3 \gamma}\>.
\end{eqnarray}
Whereas the relative amounts of $K_S K_S$ and $K_L K_L$ states are fixed
by general symmetry considerations to be equal
\begin{equation}
   \frac{\Gamma(K_S \pi^- K_S)}{\Gamma(K_L \pi^- K_L)} = 1\>,
\end{equation}
the ratio $R$ of $K_S K_S$ and $K_S K_L$ states
\begin{equation}
   R = \frac{\Gamma(K_S \pi^- K_S)}{\Gamma(K_S \pi^- K_L)}\>,
\end{equation}
is model dependent and will be discussed in the next section.
%
\section{Numerical results}
\label{numeric}
After having fixed  our model for the form factors, we next
present numerical results for the
hadronic decay widths $\Gamma{(abc)}$ normalized to the leptonic
width $\Gamma_e$
and for the branching ratios
in Tab.~\ref{tab3}.
To calculate the branching ratios, we use the theoretical prediction
for $\Gamma_e/ \Gamma_{\mbox{tot}}=17.8 \%$ based on the experimental
values for the tau mass
$m_\tau = 1.7771 \, \mbox{GeV}$ and lifetime
$\tau_{\tau} = 291.6 \, \mbox{fs}$ \protect\cite{taulifetime}, rather
than using the experimental branching ratio.

The decay rate for $\tau$ decays into three mesons can be
calculated from
\begin{equation}
\hspace{-1cm}
\Gamma(\tau\rightarrow 3h)=
           \frac{G^{2}}{12m_\tau}
\bigl(^{\cos\theta_{c}}_{\sin\theta_{c}}\bigr)^2
\frac{1}{(4\pi)^5}
\int \frac{dQ^{2}}{Q^4} ds_1 ds_2 \,(m_\tau^{2}-Q^{2})^{2}\,
          \left( 1+\frac{2Q^2}{m_\tau^2}\right)\,
           \,\,(W_{A}+W_{B})
\label{rate}
\end{equation}
where
\begin{eqnarray}  \hspace{3mm}
W_{A}  &=&   \hspace{3mm}(x_{1}^{2}+x_{3}^{2})\,|F_{1}|^{2}
           +(x_{2}^{2}+x_{3}^{2})\,|F_{2}|^{2}
           +2(x_{1}x_{2}-x_{3}^{2})\,\mbox{Re}\left(F_{1}F^{\ast}_{2}\right)
                                    \\[5mm]
W_{B}  &=& \hspace{3mm} x_{4}^{2}|F_{3}|^{2}
\label{decayrate}
\end{eqnarray}
The variables $x_i$ are defined by
$
x_{1}= V_{1}^{x}=q_{1}^{x}-q_{3}^{x},\,
x_{2}= V_{2}^{x}=q_{2}^{x}-q_{3}^{x},\,
x_{3}= V_{1}^{y}=q_{1}^{y}=-q_{2}^{y},\,
x_{4}= V_{3}^{z}=\sqrt{Q^{2}}x_{3}q_{3}^{x},\,
$
where $q_i^{x}$ ($q_i^{y}$) denotes
the $x$ ($y$) component of the momentum of
meson $i$ in the hadronic rest frame as introduced in \cite{KueMi1,KueMi2}.
They can easily be expressed in terms of $s_1$, $s_2$ and $s_3$
\cite{KueMi1}.
Eq.~(\ref{rate}) shows that there is no interference
between the axial vector contributions ($F_1, F_2$) and the
vector current contribution  ($F_3$) in the total decay
width.

Our numerical results for the normalized decay widths $\Gamma(abc)/\Gamma_e$
and the branching ratios ${\cal B}(abc)$ for our preferred parameter
choices are given in Tab.~\ref{tab3} for the various decay channels
$abc$.
To get a feeling for the numerical  importance of the vector current
({\it i.e.} the ``anomaly''),
we list its contribution to the decay width in column 3 of Tab.~\ref{tab3}.
For comparison, we have also  listed the available experimental data in
column 5 of Tab.~\ref{tab3}.

Our numerical results for ${\cal B}(K_S\pi^-K_S), {\cal B}(K^-\pi^-\pi^+)$
and ${\cal B}(\pi^-\overline{K^0}\pi^0)$ appear to be considerably higher
than the experimental results, whereas the other predictions
agree fairly well.

Our results for the $K^-\pi^-K^+$ final state do not agree with
the results in \cite{Gom90}, where the contribution
of the axial-vector channel amounts to less than 10\%
to the decay rate in this channel.
In fact, our predictions for the axial-vector contribution is
about 60\%. This result is fairly insensitive towards the details of
the $K^*$ parameterization (see Tab.~\ref{tab3},\ref{tab5}%
\ and \ref{tab6}, as discussed below).
It is however sensitive towards the $A_1$ parameters.
Use of  $\Gamma_{A_1}=0.599$ GeV in Eq.~(\ref{a1})
reduces the axial-vector contribution to about 48\%, which
is still considerably larger than the value in \cite{Gom90}.

Let us now comment on the sign of  parameter $\xi$
in the parameterization of the strange axial resonances $K_1$
 in Eq.~(\ref{xidef}),
which is only determined up to a sign ambiguity.
Our predictions for the Cabibbo suppressed decays
in Tab.~\ref{tab3} are obtained for $\xi=+0.33$.
For comparison, Tab.~\ref{tab4} shows the results for
$\xi=-0.33$.
The choice $\xi=+0.33$ is clearly preferred by the
experimental data and we therefore believe that this is
the physical choice.

Next we discuss our choice of the two body vector resonance
with strangeness, {\it i.e.} the parameterization of
$T_{K^\star}^{(1)}$ in Eq.~(\ref{betakst}).
We have used
a ${K^\star}'(1410)$ contribution in  $T_{K^\star}^{(1)}$
with a  strength of $\beta_{K^\star}=-0.135$
relative to the ${K^\star}(892)$,
as determined from $\tau \to K^\star \nu_\tau$ (see Sec.~II).
Use of $\beta_{K^\star} = -0.11$, which is also consistent
with the $\tau\rightarrow K^\star \nu_\tau$ decay rate leads to the results
in Tab.~\ref{tab5}. The results are fairly close to the numbers
presented in Tab.~\ref{tab3}.
Use of $\beta_{K^\star} = 0$ as in \cite{Dec93}
leads to the results in Tab.~\ref{tab6},
which overall agree better with the experimental results than those
obtained with $\beta_{K^\star}=-0.135$.

At this point a few comments are in order.
Firstly, by fixing the coupling constants from the chiral limit, we
effectively assume exact $SU(3)$ flavour symmetry for the coupling
constants.
The most important effects of flavour symmetry breaking have been taken
into account by using the physical masses and decay widths of the mesons
in the propagators and in the phase space. However, one could expect
additional explicit flavour symmetry breaking in the couplings.
The approximation made may
lead to an error of about $10$--$30 \%$ on the matrix
element level. Thus we consider the agreement between theory and experiment in
Tab.~\ref{tab3} as reasonable. Note that this approximation will
mainly lead to an overall normalization error of the form
factors rather than to a modification of their momentum dependence in the
relevant physical region.

Secondly, we remind the reader that a change of $\beta_{K^\star}$
has two
different effects. It changes the normalization of the $K^\star(892)$
contribution, which is proportional
to $1/(1+\beta_{K^\star})^2$ in the rate,
and it
changes the size of the $K^\star(1410)$ contribution, which is however
strongly phase space suppressed.
The main effect of choosing a different $\beta_{K^\star}$ is to
change the normalization, which may well compensate the error being made
by deriving the coupling constants using flavour symmetry.

Thirdly, we wish to emphasize that in spite of this normalization
uncertainty of the matrix elements, our parameterizations are by no means
arbitrary. We believe that they give a very good description of the
resonance substructures which determine the decay rates
as well as differential distributions and structure functions.

Our choice for $\Gamma_{A_1}=0.475$ GeV was already discussed before
[see the discussion after Eq.~(\ref{a1})].
An $A_1$ width of $\Gamma_{A_1}=0.599$ GeV decreases
the branching fractions for the  $K\pi K$ decay modes in Tab.~\ref{tab3}
by about 15\%, which is of course entirely due to a decrease of the
axial-vector contribution (see Tab.~\ref{tab1}).

As already mentioned in Sec.~\ref{neutral}, the decay rates
for $K_S\pi^-K_S$ and $K_L\pi^-K_L$ are the same. This follows
immediately from Eqs.~(\ref{klsdef},\ref{k00}) and is a model
independent statement.
On the other hand, the ratio $R$
\begin{equation}
   R = \frac{\Gamma(K_S \pi^- K_S)}{\Gamma(K_S \pi^- K_L)}\>,
\end{equation}
is model dependent and we obtain from Tab.~\ref{tab3}
\begin{equation}
R = 0.48
\end{equation}
This number is fairly insensitive towards a variation of the
$K^*$ resonance parameters.
However, use of  $\Gamma_{A_1}=.599$ GeV in Eq.~(\ref{a1}) would
lead to $R=0.42$.

We would like to point out that a study of angular correlations
of the hadronic system  allows for much more detailed studies
of the hadronic charged current (including the details of the two and
three body resonance parameters) than it is possible by
rate measurements alone.
Of particular interest is the angular distribution of the three
mesons in the three meson rest frame.
The distribution of the normal on the hadronic plane
with respect to $\vec{n}_L$ (the direction of the laboratory
as seen from the hadronic rest frame)  allows for a model independent
separation of the axial-vector  and the vector current  contribution,
{\it <i.e.} the structure functions $W_A(Q^2,s_1,s_2,s_3)$ and
$W_B(Q^2,s_1,s_2,s_3)$ in Eq.~(\ref{rate})
can be determined separately
even without reconstructing the $\tau$ rest frame \cite{KueMi1}.
An experimental analysis of the $Q^2$ and $s_i$ distributions
of these structure functions would clearly help to
test the parameterizations in Tab.~\ref{tab1} and Tab.~\ref{tab2}
unambiguously.
More general distributions like
the rotation of the mesons around the normal,
allow for even more detailed  studies of the
hadronic matrix element \cite{KueMi1}.
$Q^2$ distribution for  structure functions for
$K^-\pi^-K^+,\,K^-\pi^-\pi^+$ and $\eta\pi^-\pi^+$ final
states have already  been presented
in \cite{re} based on the parameterization in \cite{Dec93}.
We will study the  $Q^2$ and the full Dalitz plot distributions
for the decay rates and the two structure functions $W_A$ and $W_B$
based on the model in this paper in a future publication.

\section{Summary and Conclusions}
We have discussed tau decays into final states with one or two kaons.
The decays into $\pi K$ are dominated by the $K^\star$
resonance and therefore allow for a determination of the parameters of
the $K^\star$ propagator.
The experimental branching ratio can be used to
obtain a rough estimate of the $K^\star(1410)$
contribution, but we would
like to urge for a detailed study of the invariant mass distribution of
the hadronic system in order to measure
the strength of its contribution.
The decay into two kaons is predicted in agreement with experimental
data assuming dominance by the high mass tail of the $\rho$.

The three meson final states $K \pi K$ and $\pi \pi K$ allow for a much
more involved resonance substructure. We have extensively reanalyzed
these on the basis of the model of \cite{Dec93}. Our final results for
the branching ratios with our preferred parameter choices have been given
in Tab.~\ref{tab3}. They compare reasonably with experimental data,
but are rather sensitive to parameters such as $\beta_{K^\star}$.

In the case of two neutral kaons,
we have expressed the matrix elements in
terms of the $K_S$ and $K_L$ states
and given a prediction for the model dependent ratio of the
$K_S \pi^- K_S$ and $K_S \pi^- K_L$ final states.

\acknowledgments
We would like to acknowledge the contribution of Patrick Heiliger,
who started this study in collaboration with Roger Decker.
We thank  Jim  Smith and Jon Urheim for very
helpful  discussions.
The work of E. M. was supported in part
by the U. S. Department of Energy under
Grant No. DE-FG02-95ER40896.  Further support was provided by the
University of Wisconsin Research Committee, with funds granted by the
Wisconsin Alumni Research Foundation.
The work of M. F. has been supported by the HCM
program under EC contract number CHRX-CT920026.

A fortran code for the three meson matrix elements is available
from the authors [mirkes@phenom.physics.wisc.edu].
The code allows for a straightforward
implementation in the TAUOLA Monte Carlo program.

%
%
%
%


%
%
%
%
\begin{table}
\caption{Parameterization of the form factors
$F_1$ and $F_2$ in Eqs.~(\protect\ref{f1},\protect\ref{f2})
for the matrix elements of the weak axial-vector
current for the various channels.}
\label{tab1}
$$
\begin{array}{c@{\quad}c@{\quad}c@{\quad}c}
\hline \hline \\
\begin{array}{c}
\mbox{channel} \\\mbox{(abc)}
\end{array} &
A^{(abc)} & G_1^{(abc)}(Q^2,s_2,s_3) &
G_2^{(abc)}(Q^2,s_1,s_3)
\\ \\
\hline
\\
K^- \pi^- K^+ &
\frac{- \cos \theta_c}{2} &
\mbox{BW}_{A_1}(Q^2) T_\rho^{(1)}(s_2) &
\mbox{BW}_{A_1}(Q^2) T_{K^\star}^{(1)} (s_1)
\\ \\
K^0 \pi^- \overline{K^0} &
\frac{- \cos \theta_c}{2} &
\mbox{BW}_{A_1}(Q^2) T_\rho^{(1)}(s_2) &
\mbox{BW}_{A_1}(Q^2) T_{K^\star}^{(1)} (s_1)
\\ \\
K_S \pi^- K_S &
\frac{- \cos \theta_c}{4} &
\mbox{BW}_{A_1}(Q^2) T_{K^\star}^{(1)}(s_3) &
-\mbox{BW}_{A_1}(Q^2)[ T_{K^\star}^{(1)} (s_1)
+ T_{K^\star}^{(1)}(s_3) ]
\\ \\
K_S \pi^- K_L &
\frac{- \cos \theta_c}{4} &
\mbox{BW}_{A_1}(Q^2)
   [ 2 T_\rho^{(1)}(s_2) + T_{K^\star}^{(1)}(s_3)] &
\mbox{BW}_{A_1}(Q^2)[ T_{K^\star}^{(1)} (s_1)
- T_{K^\star}^{(1)}(s_3) ]
\\ \\
K^- \pi^0 K^0 &
\frac{3 \cos \theta_c}{2 \sqrt{2}} &
\!\!\!\!\mbox{BW}_{A_1} (Q^2) \left[ \frac{2}{3} T_\rho^{(1)}(s_2)
+ \frac{1}{3} T_{K^\star}^{(1)}(s_3) \right] &
\!\!\!\!\frac{1}{3}\mbox{BW}_{A_1} (Q^2) \left[  T_{K^\star}^{(1)}(s_1)
-  T_{K^\star}^{(1)}(s_3) \right]
\\ \\ \hline \\
\pi^0 \pi^0 K^- &
\frac{\sin \theta_c}{4} &
T_{K_1}^{(a)} (Q^2) T_{K^\star}^{(1)}(s_2) &
T_{K_1}^{(a)} (Q^2) T_{K^\star}^{(1)}(s_1)
\\ \\
K^- \pi^- \pi^+ &
\frac{- \sin \theta_c}{2} &
T_{K_1}^{(a)} (Q^2) T_{K^\star}^{(1)}(s_2) &
T_{K_1}^{(b)} (Q^2) T_{\rho}^{(1)}(s_1)
\\ \\
\pi^- \overline{K^0} \pi^ 0 &
\frac{3 \sin \theta_c}{2 \sqrt{2}} &
\begin{array}{l}
\,\,\,\,\frac{2}{3} T_{K_1}^{(b)} (Q^2) T_\rho^{(1)} (s_2) \\[1ex]
+ \frac{1}{3} T_{K_1}^{(a)} (Q^2) T_{K^\star}^{(1)}(s_3)
\end{array} &
\frac{1}{3} T_{K_1}^{(a)} (Q^2) \left[ T_{K^\star}^{(1)}(s_1)
 - T_{K^\star}^{(1)}(s_3) \right]
\\ \\ \hline \hline
\end{array}
$$
\end{table}


\begin{table}
\caption{Parameterization of the form factor
$F_3$ in Eq.~(\protect\ref{f3})
for the matrix elements of the weak vector current
for the various channels.}
\label{tab2}
$$
\begin{array}{c@{\quad}c@{\quad}c}
\hline \hline \\
\mbox{channel (abc)} & A^{(abc)} &
G_3^{(abc)}(Q^2,s_1,s_2,s_3)
\\ \\
\hline
\\
K^- \pi^- K^+ &
- \cos \theta_c &
T_\rho^{(2)}(Q^2) (\sqrt{2} - 1) \left[ \sqrt{2} T_\omega(s_2)
+ T_{K^\star}^{(1)}(s_1) \right]
\\ \\
K^0 \pi^- \overline{K^0} &
\cos \theta_c &
T_\rho^{(2)}(Q^2) (\sqrt{2} - 1) \left[ \sqrt{2} T_\omega(s_2)
+ T_{K^\star}^{(1)}(s_1) \right]
\\ \\
K_S \pi^- K_S &
\frac{- \cos \theta_c}{2} &
T_\rho^{(2)}(Q^2) (\sqrt{2} - 1) \left[  T_{K^\star}^{(1)}(s_1)
- T_{K^\star}^{(1)}(s_3) \right]
\\ \\
K_S \pi^- K_L &
\frac{ \cos \theta_c}{2} &
T_\rho^{(2)}(Q^2) (\sqrt{2} - 1) \left[ 2 \sqrt{2} T_\omega(s_2)
+ T_{K^\star}^{(1)}(s_1)
+ T_{K^\star}^{(1)}(s_3) \right]
\\ \\
K^- \pi^0 K^0 &
\frac{- \cos \theta_c}{\sqrt{2}} &
T_\rho^{(2)}(Q^2) (\sqrt{2} - 1) \left[  T_{K^\star}^{(1)}(s_3)
- T_{K^\star}^{(1)}(s_1) \right]
\\ \\ \hline \\
\pi^0 \pi^0 K^- &
\sin \theta_c &
\frac{1}{4} T_{K^\star}^{(2)}(Q^2) \left[ T_{K^\star}^{(1)}(s_1)
- T_{K^\star}^{(1)}(s_2) \right]
\\ \\
K^- \pi^- \pi^+ &
 \sin \theta_c &
\frac{1}{2} T_{K^\star}^{(2)}(Q^2) \left[ T_\rho^{(1)}(s_1)
+  T_{K^\star}^{(1)}(s_2) \right]
\\ \\
\pi^- \overline{K^0} \pi^ 0 &
 \sqrt{2} \sin \theta_c &
\frac{1}{4} T_{K^\star}^{(2)}(Q^2) \left[2 T_\rho^{(1)}(s_2)
+ T_{K^\star}^{(1)}(s_1)
+ T_{K^\star}^{(1)}(s_3) \right]
\\ \\ \hline \hline
\end{array}
$$
\end{table}


\begin{table}
\caption{Predictions for the normalized decay widths
$\Gamma(abc)/\Gamma_e$ and the branching
ratios ${\cal B}(abc)$ for the various channels
with $\beta_{K^\star}=-0.135$ in Eq.~(\protect\ref{betakst}).
The contribution from the vector current is listed in column 3 and
available experimental data are listed in column 5.
The later are taken from \protect\cite{Hel94},
except for
$K_S \pi^- K_S$, where we quote the value for $K_S h K_S$ given in
\protect\cite{Smi94}.
}.
\label{tab3}
$$
\begin{array}{c@{\quad}c@{\quad}c@{\quad}c@{\quad}c}
\hline \hline \\
\mbox{channel (abc)} &
\left( \frac{\Gamma{(abc)}}{\Gamma_e} \right)^{(pred.)} &
\left( \frac{\Gamma{(abc)}}{\Gamma_e} \right)^{(pred.)}_{anomaly} &
\left({\cal B}{(abc)} \right)^{(pred.)} [\%] &
\left({\cal B}{(abc)} \right)^{(expt.)} [\%]
 \\ \\
\hline
\\[-4mm]
K^- \pi^- K^+            & 0.011   & 0.0045  & 0.20  &  0.20 \pm 0.07  \\
K^0 \pi^- \overline{K^0} & 0.011   & 0.0045  & 0.20  &                 \\
K_S \pi^- K_S            & 0.0027  & 0.0008  & 0.048 &  0.021\pm 0.006 \\
K_S \pi^- K_L            & 0.0058  & 0.0029  & 0.10  &                 \\
K^- \pi^0 K^0            & 0.0090  & 0.0032  & 0.16  &  0.12 \pm 0.04  \\[2mm]
\hline \\[-4mm]
\pi^0 \pi^0 K^-          & 0.0080  & 0.0007  & 0.14  &  0.09 \pm 0.03  \\
K^-   \pi^-\pi^+         & 0.043   & 0.0043  & 0.77  &  0.40 \pm 0.09  \\
\pi^-\overline{K^0}\pi^0 & 0.054   & 0.0058  & 0.96  &  0.41 \pm 0.07  \\[2mm]
\hline \hline
\end{array}
$$
\end{table}


\begin{table}
\caption{
Same as Tab.~\protect\ref{tab3} for the Cabibbo suppressed decays with
$\xi=-0.33$ in Eq.~(\protect\ref{xidef}).
  }
\label{tab4}
$$
\begin{array}{c@{\quad}c@{\quad}c@{\quad}c@{\quad}c}
\hline \hline \\
\mbox{channel (abc)} &
\left( \frac{\Gamma{(abc)}}{\Gamma_e} \right)^{(pred.)} &
\left( \frac{\Gamma{(abc)}}{\Gamma_e} \right)^{(pred.)}_{anomaly} &
\left({\cal B}{(abc)} \right)^{(pred.)} [\%] &
\left({\cal B}{(abc)} \right)^{(expt.)} [\%]
 \\ \\
\hline
\\[-4mm]
\pi^0 \pi^0 K^-          & 0.017   & 0.0007  & 0.30  &  0.09 \pm 0.03  \\
K^- \pi^- \pi^+          & 0.077   & 0.0043  & 1.37  &  0.40 \pm 0.09  \\
\pi^-\overline{K^0}\pi^0 & 0.087   & 0.0058  & 1.55  &  0.41 \pm 0.07  \\[2mm]
\hline \hline
\end{array}
$$
\end{table}


\begin{table}
\caption{
Same as Tab.~\protect\ref{tab3}
but with $\beta_{K^\star}=-0.11$ in Eq.~(\protect\ref{betakst}).
   }.
\label{tab5}
$$
\begin{array}{c@{\quad}c@{\quad}c@{\quad}c@{\quad}c}
\hline \hline \\
\mbox{channel (abc)} &
\left( \frac{\Gamma{(abc)}}{\Gamma_e} \right)^{(pred.)} &
\left( \frac{\Gamma{(abc)}}{\Gamma_e} \right)^{(pred.)}_{anomaly} &
\left({\cal B}{(abc)} \right)^{(pred.)} [\%] &
\left({\cal B}{(abc)} \right)^{(expt.)} [\%]
 \\ \\
\hline
\\[-4mm]
K^- \pi^- K^+            & 0.0106  & 0.0042  & 0.19  &  0.20 \pm 0.07  \\
K^0 \pi^- \overline{K^0} & 0.0106  & 0.0042  & 0.19  &                 \\
K_S \pi^- K_S            & 0.0026  & 0.0007  & 0.046 &  0.021\pm 0.006 \\
K_S \pi^- K_L            & 0.0055  & 0.0027  & 0.098 &                 \\
K^- \pi^0 K^0            & 0.0085  & 0.0030  & 0.15  &  0.12 \pm 0.04  \\[2mm]
\hline \\[-4mm]
\pi^0 \pi^0 K^-          & 0.0076  & 0.0007  & 0.14  &  0.09 \pm 0.03  \\
K^- \pi^- \pi^+          & 0.041   & 0.0041  & 0.74  &  0.40 \pm 0.09  \\
\pi^-\overline{K^0}\pi^0 & 0.052   & 0.0056  & 0.93  &  0.41 \pm 0.07  \\[2mm]
\hline \hline
\end{array}
$$
\end{table}


\begin{table}
\caption{
Same as Tab.~\protect\ref{tab3}
but with $\beta_{K^\star}=0$ in Eq.~(\protect\ref{betakst}).
}
\label{tab6}
$$
\begin{array}{c@{\quad}c@{\quad}c@{\quad}c@{\quad}c}
\hline \hline \\
\mbox{channel (abc)} &
\left( \frac{\Gamma{(abc)}}{\Gamma_e} \right)^{(pred.)} &
\left( \frac{\Gamma{(abc)}}{\Gamma_e} \right)^{(pred.)}_{anomaly} &
\left({\cal B}{(abc)} \right)^{(pred.)} [\%] &
\left({\cal B}{(abc)} \right)^{(expt.)} [\%]
 \\ \\
\hline
\\[-4mm]
K^- \pi^- K^+            & 0.0084  & 0.0034  & 0.15  &  0.20 \pm 0.07  \\
K^0 \pi^- \overline{K^0} & 0.0084  & 0.0034  & 0.15  &                 \\
K_S \pi^- K_S            & 0.0020  & 0.0006  & 0.036 &  0.021\pm 0.006 \\
K_S \pi^- K_L            & 0.0044  & 0.0022  & 0.078 &                 \\
K^- \pi^0 K^0            & 0.0067  & 0.0024  & 0.12  &  0.12 \pm 0.04  \\[2mm]
\hline \\[-4mm]
\pi^0 \pi^0 K^-          & 0.0060  & 0.0005  & 0.11  &  0.09 \pm 0.03  \\
K^- \pi^- \pi^+          & 0.035   & 0.0035  & 0.62  &  0.40 \pm 0.09  \\
\pi^-\overline{K^0}\pi^0 & 0.045   & 0.0048  & 0.81  &  0.41 \pm 0.07  \\[2mm]
\hline \hline
\end{array}
$$
\end{table}
\end{document}